\documentclass[aps]{revtex4}
\usepackage{graphicx}
\usepackage{amsmath}

\begin{document}
\title{Monopoles in Lattice QCD with Abelian Projection as 
Quantum Monopoles}
\author{V. Dzhunushaliev}
\email{dzhun@hotmail.kg}
\affiliation{Dept. of Phys. and Microel. Engineer., 
Kyrgyz-Russian Slavic University,
Bishkek, 720000, Kyrgyz Republic}
\author{D. Singleton}
\email{dougs@csufresno.edu}
\affiliation{Physics Dept., CSU Fresno, 2345 East San Ramon Ave.
M/S 37 Fresno, CA 93740-8031, USA}

\begin{abstract}
Within the context of the Abelian Projection of QCD monopole-like 
quantum excitations of gauge fields are studied. We start with
certain classical solutions, of the SU(2) Yang-Mills field
equations, which are not monopole-like and
whose energy density diverges as $r \rightarrow \infty$.
These divergent classical solutions are then quantized using a
modified version of Heisenberg's quantization technique for strongly
interacting, nonlinear fields. The modified Heisenberg quantization
technique leads to a system of equations with mixed quantum and
classical degrees of freedom. By applying a Feynman path integration
over the quantum degrees of freedom the quantum-averaged solution
gives a nondivergent, monopole-like configuration after Abelian
Projection.
\end{abstract}

%\pacs{12.38.Lg}

\maketitle

\section{Introduction}
Monopoles have been studied within both
Abelian \cite{dirac} and non-Abelian \cite{thooft} gauge
theories. Usually one starts with a
classical monopole configuration ({\it i.e.} having a
magnetic field which becomes Coulombic as $r \rightarrow
\infty$) and then consider the quantum corrections
to the system. One difficulty with this approach is
that the coupling strength of monopole configurations
is large due to the Dirac quantization condition \cite{dirac}
which requires that there be an inverse relationship between
electric and magnetic couplings. A perturbatively small
electric coupling requires a non-perturbatively large
magnetic coupling. The perturbative quantization techniques
do not work well with monopoles for the same reason that they
have trouble with QCD in the low energy regime : the couplings
are non-perturbatively large.
\par 
Numerical simulations in lattice QCD show that the configurations 
of gauge fields with monopoles dominate in the path integral, 
\emph{i.e.} monopole configurations are important
contributors to the QCD string tension.
Thus the expectation value of any physical quantity 
$\langle \Upsilon \rangle$
in non-Abelian gauge theory can be
accurately calculated (see for review Ref. \cite{pol}) 
in terms of the monopole currents
extracted from the projected Abelian fields. Magnetic monopoles
necessarily emerge as relevant degrees of freedom in the Abelian 
Projections of lattice gluodynamics. In this paper we present 
a specific picture of how such magnetic monopole configurations
can emerge and dominate the QCD path integral.
We start with spherically  symmetric classical solutions
to the SU(2) field equations. These solutions have a divergent
energy density. However in constrast to the Wu-Yang monopole
solutions, which are divergent due to fields that blow up at
$r=0$, our solutions are divergent due to fields which blow up
at $r=\infty$. These divergent solutions are parameterized
by two ansatz functions. One of the ansatz functions smoothly and
monotonically diverges at $\infty$ while the other functions
strongly oscillates. Applying a modified variant of
a non-perturbative quantization method originally used by
Heisenberg \cite{hs1} \cite{hs3}, we find that the quantized version
of our divergent classical solution becomes well behaved: the
divergent ansatz function now goes to $0$ at $\infty$, and the
rapid ocsillations of the other ansatz function are ``smoothed''
out. Rather than quantizing an SU(2) field configuration
which is monopole-like already at the classical level,
we quantized (using Heisenberg's method) a non-monopole
configuration, which has fields and an energy density which diverge
at spacial infinity ($r \rightarrow \infty$). After
quantization this classically divergent solution becomes
physically well behaved, and its
asymptotic magnetic field becomes monopole-like. This
may indicate that monopoles are inherently
quantum objects: rather than quantizing
a configuration which is monopole-like at the classical level,
real monopoles might arise as a consequence of quantizing
non-monopole classical solutions.

\section{Classical SU(2) solution}

In this section we will briefly review the classical,
spherically symmetric, SU(2) Yang-Mills theory solutions 
to which we will apply the modified Heisenberg quantization
method. We begin with the following ansatz for pure SU(2) Yang-Mills
theory
\begin{eqnarray}
    A^a_0 & = & \frac{x^a}{r^2} g(r),
\label{s1-1}\\
    A^a_i & = & \epsilon_{aij}\frac{x^j}{r^2} 
    \left[ 1 - f(r) \right]
\label{s1-2},
\end{eqnarray}
here $a=1,2,3$ is a color index; the Latin indices $i,j,k,l=1,2,3$
are space indices. This is the Wu-Yang ansatz \cite{wu}.
\par
\begin{figure}[htb]
\begin{center}
\fbox{
\includegraphics[height=5cm,width=5cm]{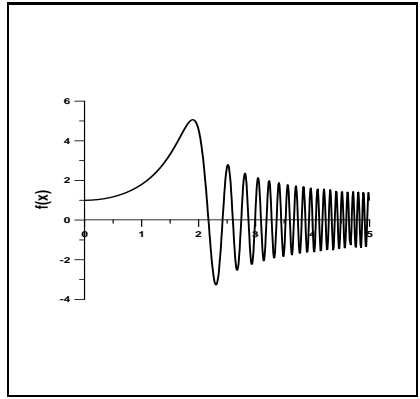}
}
\caption{Function $f(x)$}
\label{fig1}
\end{center}
\end{figure}
\begin{figure}[htb]
\begin{center}
\fbox{
\includegraphics[height=5cm,width=5cm]{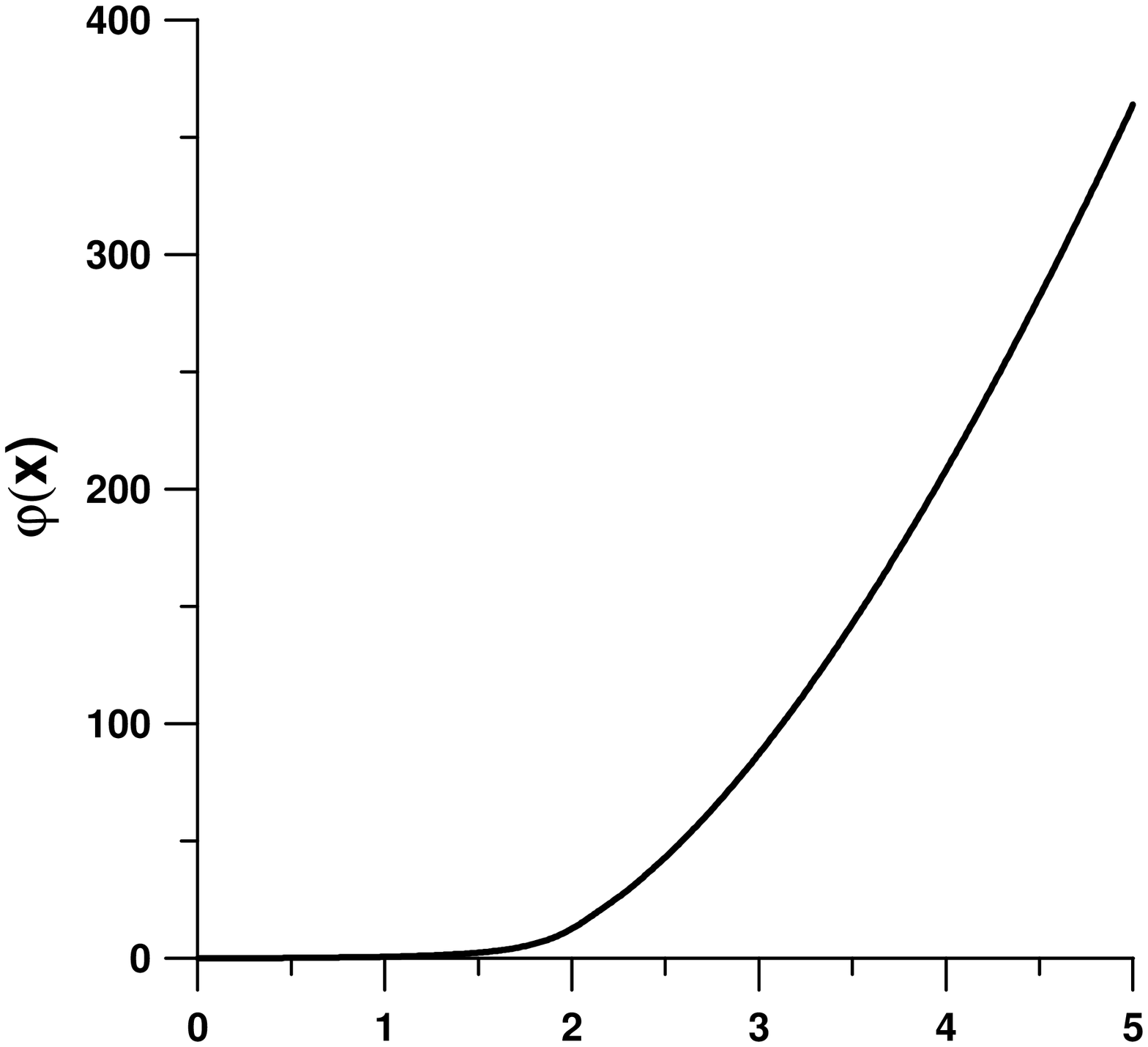}
}
\caption{Function $g(x)$}
\label{fig2}
\end{center}
\end{figure}

Under this ansatz the Yang - Mills equations
($D^{\mu} {F^a}_{\mu \nu}= 0$) become
\begin{eqnarray}
    r^2 f''& = & f^3 - f - fg^2,
\label{s1-3}\\
    r^2 g''& = & 2g f^2
\label{s1-4}
\end{eqnarray}
where the primes indicate differentiation with respect to $r$.
In the asymptotic limit $r \rightarrow \infty$ the general solution 
to Eqs. \eqref{s1-3}, \eqref{s1-4} approaches the form
\begin{eqnarray}
    f(x) & \approx & A \sin \left (x^{\alpha } + \phi _0\right ),
\label{s1-5}\\
    g(x) & \approx & \alpha  x^{ \alpha } +
\frac{\alpha -1 }{4}\frac{\cos {\left (2x^{\alpha} + 2\phi _0 \right )}}
{x^{\alpha}},
\label{s1-6}\\
    A^2 & = & \alpha(\alpha - 1)
\label{s1-7}
\end{eqnarray}
where $x=r/r_0$ is a dimensionless radius and $r_0, \phi _0$, 
and $A$ are constants. The plot of these functions are presented 
on the Fig.\ref{fig1}, \ref{fig2}. 

The numerical study of Eqs. \eqref{s1-3}, \eqref{s1-4} showed
that the approximate form of the
solution in Eqs. \eqref{s1-5} - \eqref{s1-7} was a good representation
even for $r$ close to the origin. These solutions can be compared
with the certain confining potentials used in phenomenological
studies of quarkonia bound states \cite{eich}. It is possible
to find analytic solutions with increasing
gauge potentials, similar to the above numerical solution for
the ansatz $g(x)$, if one couples the Yang-Mills fields
to a scalar field \cite{yos}. However, it is the 
increasing of $g(r)$ that leads to the field energy and action
of this solution diverging as $r \rightarrow \infty$.
\par
The ``magnetic'' ($H_i ^a = \epsilon_{ijk}F^a _{jk}$) and ``electric''
($E_i ^a = F^a _{0i}$) fields associated with this solution
can be found from the non-Abelian gauge potentials, $A_{\mu} ^a$,
and have the following behavior
\begin{alignat}{3}
H^a _r & \propto & \frac{f^2-1}{r^2} ,
& \qquad H^a_{\varphi}  \propto  f' ,
& \qquad H^a_{\theta}  \propto  f' ,
\label{s1-8}\\
E^a_r & \propto & \frac{rg' - g}{r^2},
& \qquad E^a_{\varphi} \propto  \frac{fg}{r},
& \qquad E^a_{\theta}  \propto  \frac{fg}{r}.
\label{s1-9}
\end{alignat}

\section{Extraction of the off-diagonal components of 
gauge potential}

Our SU(2) gauge potential \eqref{s1-1} \eqref{s1-2} can be 
written as
\begin{eqnarray}
  A^a_\theta & = & \left[1 - f(r)\right] 
  \left\{
    \sin\varphi ; -\cos\varphi ; 0
  \right\},
\label{sec3-1}\\
  A^q_\varphi & = & \left[1 - f(r)\right] \sin\theta
  \left\{
    \cos\theta \cos\varphi ; \cos\theta \sin\varphi; 
    -\sin\theta
  \right\},
\label{sec3-2}\\
  A^a_t & = & \frac{g(r)}{r}
  \left\{
    \sin\theta \cos\varphi ; \sin\theta \sin\varphi ; \cos\theta
  \right\}.
\label{sec3-3}  
\end{eqnarray}
The gauge transformation 
\begin{equation}
  {A'}^a_\mu = S^{-1} A^a_\mu S - 
  i \left( \partial_\mu S^{-1} \right) S
\label{sec3-4}
\end{equation}
with the matrix 
\begin{equation}
  S = \begin{pmatrix}
  \cos\frac{\theta}{2} & - e^{-i\varphi} \sin\frac{\theta}{2}\\
  e^{i\varphi} \sin\frac{\theta}{2} & \cos\frac{\theta}{2} 
  \end{pmatrix}
\label{sec3-5}
\end{equation}
leads to the following result 
\begin{eqnarray}
  A'_\theta & = & f(r)
  \begin{pmatrix}
    0 & - ie^{-i\varphi} \\
    i e^{i\varphi} & 0 
  \end{pmatrix},
\label{sec3-6}\\
  A'_\varphi & = & \left( \cos\theta - 1 \right)
  \begin{pmatrix}
    1 & 0 \\
    0 & -1 
  \end{pmatrix} - f(r) \sin\theta 
  \begin{pmatrix}
    0 & e^{-i\varphi} \\
    e^{i\varphi} & 0 
  \end{pmatrix},
\label{sec3-7}\\
  A'_t & = & \frac{g(r)}{r}
  \begin{pmatrix}
    1 & 0 \\
    0 & -1 
  \end{pmatrix}.
\label{sec3-8}  
\end{eqnarray}
In this form we see that if $f(r) \rightarrow 0$ as $r \rightarrow \infty$
then only the first term in Eq. \eqref{sec3-7}
is relevant. This first term is in the U(1) subgroup of the SU(2)
gauge field, and gives rise to a magnetic monopole.
\par 
Under the assumption (originally found in ref. \cite{thooft2}) that
the U(1) diagonal components of the non-Abelian gauge field are
the most important for confinement, one tries to make a gauge choice
which minimizes or excludes the off-diagonal components of the gauge
potential ({\it i.e.} the Cartan subalgebra).
For example, in Maximal Abelian Projection \cite{kronfeld}, \cite{kondo} 
the following gauge condition is applied
\begin{equation}
  D_\mu \left( A^H \right) A^{off}_\mu = 
  \partial_\mu A^{off}_\mu + ig \left[ A^H_\mu , A^{off}_\mu \right] 
  = 0 
\label{sec3-9}
\end{equation}
where $A_\mu = A^H_\mu + A^{off}_\mu$. $A^{H}_\mu $ 
is the Abelian, diagonal pomponent of $A_\mu$ while
$A^{off}_\mu$ are the off-diagonal components of $A_\mu$.
This gauge corresponds to the minimization of the 
following functional \cite{pol} 
\begin{equation}
  R = \int \left\{
  \left[ A^1_\mu \right]^2 + 
  \left[ A^2_\mu \right]^2 \right\} d^4 x
\label{sec3-10}
\end{equation}
where $A^{off}_\mu = A^{1,2}_\mu$. Lattice calculations
(for review, see for example, \cite{pol2}) indicate that monopole-like 
gauge field configurations
are dominate in the path integral. Here we show that
one can reach this same conclusion by using a variant of Heisenberg's
quantization method on non-monopole, classical solution. 
By quantizing the non-Abelian we find that $f(r) \rightarrow 0$ as
$r \rightarrow \infty$. Thus the
quantized non-Abelian fields asymptotically appear as monopoles.

\section{Heisenberg's Non-Perturbative Quantization}

Although the confining behavior of these classical solutions
is interesting due to their similarity with certain phenomenological,
confining potentials, the infinite field energy makes
their physical importance/meaning uncertain.
One possible resolution to the
divergent field energy is if quantum effects
removed the bad long distance behavior. The difficulty is that
strongly interacting, nonlinear theories are notoriously hard to
quantize. In order to take into account the quantum effects
on these solutions we will employ a variation of the method
used by Heisenberg \cite{hs1} in attempts to
quantize the nonlinear Dirac equation. 
It can be shown \cite{dzh4} that the Gor'kov's equations (and 
consequently the Ginzburg - Landau equations) in superconductivity 
theory is the direct application of Heisenberg's idea for quantization 
of Cooper pairs in the superconductor. We will outline
the key points of the method by using the nonlinear
Dirac equation as an illustrative example.
The nonlinear spinor field equation considered by
Heisenberg had the following form :
\begin{equation}
\label{hq-1}
\gamma ^{\mu} \partial _{\mu} {\hat \psi (x)} - l^2
\Im [{\hat \psi} ({\hat {\bar \psi}}
\psi) ] = 0
\end{equation}
where $\gamma ^{\mu}$ are Dirac matrices; ${\hat \psi (x)},
{\hat {\bar \psi}}$ are the spinor field and its adjoint
respectively; $\Im [{\hat \psi} ({\hat {\bar \psi}} {\hat \psi}) ]$
is the general nonlinear spinor self interaction term which
involved three spinor fields and various combinations of $\gamma^{\mu}$'s
and/or $\gamma ^5$'s. The constant $l$ has units of length, and sets the
scale for the strength of the interaction. Next one defines $\tau$
functions as
\begin{equation}
\label{hq-2}
\tau (x_1 x_2 ... | y_1 y_2 ...) = \langle 0 | T[{\hat \psi} (x_1)
{\hat \psi} (x_2) ... {\hat \psi} ^{\ast}
(y_1) {\hat \psi} ^{\ast} (y_2) ...] | \Phi \rangle
\end{equation}
where $T$ is the time ordering operator; $| \Phi \rangle$ is a state for
the system described by Eq. \eqref{hq-1}. Applying
Eq. \eqref{hq-1} to \eqref{hq-2} we obtain the following infinite
system of equations for various $\tau $'s
\begin{eqnarray}
\label{hq-3}
l^{-2} \gamma ^{\mu} _{(r)} \frac{\partial}{\partial x^{\mu} _{(r)}}
&&\tau (x_1 ...x_n |y_1 ... y_n ) = \Im [ \tau (x_1 ... x_n x_r |
y_1 ... y_n y_r)] + \nonumber \\
&&\delta (x_r -y_1) \tau ( x_1 ... x_{r-1} x_{r+1} ... x_n |
y_2 ... y_{r-1} y_{r+1} ... y_n ) + \nonumber \\
&&\delta (x_r - y_2) \tau (x_1 ... x_{r-1} x_{r+1} ... x_n |
y_1 y_2 ... y_{r-1} y_{r+1} ... y_n ) + ...
\end{eqnarray}
Eq. \eqref{hq-3} represents one of an infinite set of coupled equations
which relate various order (given by the index $n$) of the $\tau$
functions to one another. To make some head way toward solving
this infinite set of equations Heisenberg employed the Tamm-Dankoff
method whereby he only considered $\tau$ functions up to a certain
order. This effectively turned the infinite set of coupled equations
into a finite set of coupled equations.
\par
For the SU(2) Yang-Mills theory this idea leads to the following
Yang-Mills equations for the quantized SU(2) gauge field
\begin{equation}
  D_\mu \hat F^{a\mu\nu} = 0
  \label{hq3a}
\end{equation}
here $\hat F^{a\mu\nu}$ is the field {\it operator} of the SU(2)
gauge field.
\par
One can show that Heisenberg's method
is equivalent to the Dyson-Schwinger system of equations for
small coupling constants. One can also make a comparison
between the Heisenberg method and the standard Feynman
diagram technique. With the Feynman diagram method quantum
corrections to physical
quantities are given in terms of an infinite number of higher
order, loop diagrams. In practice one takes only a finite
number of diagrams into account when calculating the quantum
correction to some physical quantity.
This standard diagrammatic method requires a small
expansion parameter (the coupling constant), and thus
does not work for strongly coupled theories. The Heisenberg
method was intended for strongly coupled, nonlinear theories,
and we will apply a variation of this method to the classical
solution discussed in the last section.

We will consider a variation of Heisenberg's quantization method
for the present non-Abelian equations by making the following
assumptions \cite{dzh2}:
\begin{enumerate}
\item
The physical degrees of freedom relevant for studying
the above classical solution (which after quantization will be 
spherically symmetric excitations in QCD vacuum) 
are given entirely by the two ansatz
functions $f, g$ appearing in Eqs. \eqref{s1-3}, \eqref{s1-4}.
No other degrees of freedom will arise through
the quantization process.
\item
From Eqs. \eqref{s1-5}, \eqref{s1-6} we see that one function
$f(r)$ is a smoothly varying function for large
$r$, while another function, $g(r)$, is strongly oscillating.
Thus we take $g(r)$ to be an almost
classical degree of freedom while
$f(r)$ is treated as a
fully quantum mechanical degree of freedom.
Naively one might expect that
only the behavior of second function
would change while the first function stayed
the same. However since both functions are interrelated
through the nonlinear nature of the field equations
we find that both functions are modified.
\end{enumerate}
To begin we replace the ansatz functions by operators ${\hat f} (x) ,
{\hat g} (x)$.
\begin{eqnarray}
\label{s3-1}
r^2 {\hat f}'' &=& {\hat f}^3 - {\hat f} - {\hat f} {\hat g} ^2 ,
\label{s3-2} \\
r^2 {\hat g}'' &=& 2 {\hat g} {\hat f}^2
\label{s3-3}
\end{eqnarray}
These equations can be seen as an approximation of
the quantized SU(2) Yang-Mills field equations \eqref{hq3a}.
Taking into account assumption (2) we let
${\hat g} \rightarrow g$ so that it becomes just
a classical function again, and replace
${\hat f}^2$ in Eq. \eqref{s3-3}
by its expectation value to arrive at
\begin{eqnarray}
r^2 {\hat f}'' &=& {\hat f}^3 - {\hat f} - {\hat f} g ^2 ,
\label{s3-4} \\
r^2 g'' &=& 2 g \langle f^2 \rangle
\label{s3-5}
\end{eqnarray}
where the expectation value $\langle {\hat f}^2 \rangle$ is taken with
respect to some quantum state $|q\rangle$ --
$\langle f^2 \rangle = \langle q| f^2 | q\rangle $.
We can average Eq. \eqref{s3-4} to get
\begin{eqnarray}
  r^2 \langle f \rangle '' & = &
  \langle  f^3 \rangle - \langle f \rangle -
  \langle f \rangle g^2 ,
  \label{s3-6}\\
  r^2 g'' & = & 2 g \langle f^2 \rangle
  \label{s3-7}
\end{eqnarray}
Eqs. \eqref{s3-6} \eqref{s3-7} are almost a closed system for
determining $\langle f \rangle$ except for the $\langle f^2 \rangle$
and $\langle f^3 \rangle$ terms. One can obtain differential equations
for these expectation values by applying $r^2 \partial
/ \partial r$ to ${\hat f}^2$ or ${\hat f}^3$ and using
Eqs. \eqref{s3-4} - \eqref{s3-5}. However the differential
equations for $\langle f^2 \rangle$ or $\langle f^3 \rangle$
would involve yet higher powers of ${\hat f}$ thus generating
an infinite number of coupled differential equations for the various
$\langle f^n \rangle$. In the next section we will use a path integral
{\em inspired} method \cite{dzh3} to cut this progression off at some finite
number of differential equations.

\section{Path integration over classical solutions}

Within the path integral method the expectation value of
some field $\Phi$ is given by
\begin{equation}
\label{av-2}
  \langle \Phi \rangle = \int  \Phi e^{i S
  \left [  \Phi \right ]} D \Phi .
\end{equation}
The classical solutions (if they exist), $\Phi _{cl}$, give the dominate
contribution to the path integral. For a single classical
solution one can approximate the path integral as
\begin{equation}
  \int e^{iS[\Phi]} D\Phi \approx A e^{iS[\Phi_{cl}]}
  \label{av-2a}
\end{equation}
where $A$ is a normalization constant. Consequently the
expectation of the field can be approximated by
\begin{equation}
  \int \Phi e^{iS[\Phi]} D\Phi \approx \Phi_{cl} .
  \label{av-3b}
\end{equation}
We are interested in the case where $\Phi$ is the gauge
potential $A^a_\mu$, in which case our approximation becomes
\begin{equation}
  \langle A^a_\mu \rangle \approx
  \int \left (\tilde A^a_\mu \right )_{\phi _0}
  e^{i S  \left [ \left (\tilde A^a_\mu
  \right )_{\phi _0} \right ]}
  D \left (\tilde A^a_\mu \right )_{\phi _0}
  \label{av-1}
\end{equation}
$(\tilde A^a_\mu)_{\phi _0}$  are the classical
solutions of the Yang - Mills equations labelled by a parameter
$\phi _0$. In the present case the classical solution with
the asymptotic form \eqref{s1-5} \eqref{s1-6} has an infinite
energy and action. When one considers the Euclidean version
of the path integral above the exponential factor in
\eqref{av-1} becomes $\exp[-S [ (\tilde A^a_\mu
)_{\phi_0}]$ which for an infinite action
would naively imply that this classical configuration
would not contribute to the path integral at all. However,
there are examples where infinite action, classical solutions
have been hypothesized to play a significant role in the
path integral. The most well known example of this is
the meron solution \cite{callan}, which has an infinite
action. Analytically the singularities of the meron
solutions can be dealt with by replacing the regions that
contain the singularities by instanton solutions. Since
instanton solutions have finite action this patched
together solution of meron plus instanton has finite action.
The drawback is that the Yang-Mills field equations are not
satisfied at the boundary where the meron and instanton
solutions are ``sewn'' together. In addition recent
lattice studies \cite{sn} \cite{negele}
have indicated that merons (or the patched meron/instanton)
do play a role in the path integral. Here we will treat
this divergence in the action in an approximate way through
a redefinition of the path integral integration measure.
We now define the path integral \eqref{av-1} 
by the following approximate way 
\begin{equation}
  \langle A^a_\mu (r) \rangle = \int A^a_\mu e^{-S[A]} DA^a_\mu  
  \approx 
  \frac{
  \int \left( \tilde A^a_\mu (r) \right)_{cl}
  e^{-i\int L\left[ \left( \tilde A^a_\mu (r) \right)_{cl}  
   \right] dV}
  d \left( \tilde A^a_\mu (r) \right)_{cl}
  }
  {\int e^{-i \int L\left[ \left( \tilde A^a_\mu (r) \right)_{cl}  
   \right] dV}
  d \left( \tilde A^a_\mu (r) \right)_{cl}
  }
\label{av-2b}
\end{equation}
where $dV$ is the 3D infinitesimal volume. 
Here our basic assumption is that in the first rough approximation 
$\langle A^a_\mu (r) \rangle$ can be approximately estimated by 
integration over classical singular solutions. Every such solution 
is labelled by $\phi_0$ and consequently we should integrate over 
phase $\phi_0$. It can be shown that the Lagrangian for the system is
\begin{equation}
  L \approx \alpha^2 (\alpha - 1) 
  \left\{
  \alpha - 1 + 2\alpha \sin^2 
  \left[ \left( \frac{r}{r_0} \right)^\alpha + \phi_0\right]
  \right\} r^{2\alpha}.
\label{av-3a}
\end{equation}
The next our assumption is that we neglect the oscillating term 
$\sin^2 [(r/r_0)^\alpha + \phi_0 ]$ and factor 
$e^{-i\int \alpha^2 (\alpha - 1)^2r^{2\alpha} dV}$ is cancelled in 
numerator and denominator. That leads to 
\begin{equation}
  \langle A^a_\mu (r) \rangle \approx 
  \frac{\int d\phi_0 \left( \tilde A^a_\mu (r) \right)_{\phi_0}}
  {2\pi}
\label{av-4a}
\end{equation}
here the measure 
$d \left( \tilde A^a_\mu (r) \right)_{cl} = d\phi_0$. Using this
and the fact that $A^a_t(r) \propto f(r)$ we calculate the
expectation value of various powers of $f(r)$
\begin{eqnarray}
  \langle f \rangle & \approx &
  \frac{1}{2\pi} \int\limits^{2\pi}_0 f_{cl} d\phi _0 =
  \frac{A}{2\pi}
  \int\limits^{2\pi}_0 \sin(x^\alpha + \phi _0) d\phi _0  = 0 ,
  \label{av-3}\\
  \langle f^2 \rangle & \approx &
  \frac{1}{2\pi} \int\limits^{2\pi}_0 f^2_{cl} d\phi _0 =
  \frac{A^2}{2\pi}  \int\limits^{2\pi}_0
  \sin^2 (x^\alpha + \phi _0) d\phi _0 =
  \frac{A^2}{2} ,
  \label{av-4}\\
  \langle f^3 \rangle & \approx &
  \frac{1}{2\pi} \int\limits^{2\pi}_0 f^3_{cl} d\phi _0 =
  \frac{A^3}{2\pi}  \int\limits^{2\pi}_0
  \sin^3 (x^\alpha + \phi _0) d\phi _0 = 0 ,
  \label{av-5}\\
  \langle f^4 \rangle & \approx &
  \frac{1}{2\pi} \int\limits^{2\pi}_0 f^4_{cl} d\phi _0 =
  \frac{A^4}{2\pi}  \int\limits^{2\pi}_0
  \sin^4 (x^\alpha + \phi _0) d\phi _0 =
  \frac{3}{8} A^4
  \label{av-6}
\end{eqnarray}
where $f_{cl}$ is the function from the Eq.\eqref{s1-5}.
The path integral {\em inspired} Eqns. \eqref{av-3} -
\eqref{av-6} are the heart of the cutoff procedure
that we wish to apply to Eqns. \eqref{s3-6}, \eqref{s3-7}.
On substituting Eqs. \eqref{av-3}, \eqref{av-4}
into Eqs. \eqref{s3-6}, \eqref{s3-7} we find that
Eq.\eqref{s3-6} is satisfied identically
and Eq.\eqref{s3-7} takes the form
\begin{equation}
  r^2 g'' = A^2g = \alpha (\alpha - 1) g ,
  \qquad \alpha >1
  \label{av-7}
\end{equation}
which has the solutions
\begin{eqnarray}
  g & = & g_0 r^\alpha ,
  \label{av-8}\\
  g & = & \frac{g_0}{r^{\alpha - 1}}
  \label{av-9}
\end{eqnarray}
where $g_0$ is some constant. The first
solution is simply the classically averaged singular solution
\eqref{s1-6} which still has the bad asymptotic divergence of the
fields and energy density. The second solution,
\eqref{av-9}, is more physically relevant since
it leads to asymptotic fields which are well behaved.
\par
The solution of Eq. \eqref{av-9} implies the following important
result: \textbf{\textit{the quantum fluctuations
of the strongly oscillating, nonlinear fields leads to an improvement
of the bad asymptotic behavior of these nonlinear fields.}}
After quantization the monotonically growing
and strongly oscillating components of the gauge potential
became asymptotically well behaved. We interpret this spherically
symmetric quantized field distribution \textbf{as a model of the
monopole excitations in QCD vacuum.} 
\par
As $r \rightarrow \infty$ we find the following SU(2) color fields
\begin{eqnarray}
  \langle H^a_r \rangle & \propto &
  \frac{\langle f^2 \rangle - 1}{r^2}
  \approx
  \frac{Q}{r^2}
  \qquad \text{with} \qquad
  Q = \frac{1}{2}\alpha (\alpha -1 ) - 1 ,
  \label{av-10}\\
  \langle H^a_{\varphi ,\theta} \rangle & \propto &
  \langle f' \rangle \approx 0 ,
  \label{av-11}\\
  \langle E^a_r \rangle & \propto &
  \langle \frac{rg' - g}{r^2} \rangle
  \approx
  -\frac{\alpha g_0}{r^{\alpha + 1}} ,
  \label{av-12}\\
  \langle E^a_{\varphi , \theta} \rangle & \propto &
  \frac{\langle f \rangle g}{r} = 0 .
  \label{av-13}
\end{eqnarray}
We can see that as $r \rightarrow \infty$
$|\langle E^a_r \rangle | \ll |\langle H^a_r \rangle |$.
In particular at infinity we find only a monopole
``magnetic'' field $H^a_r \approx Q/r^2$ with a ``magnetic''
charge $Q$. This result can be summarized as:
\textbf{\textit{the approximate quantization of the SU(2) gauge field
(by averaging over the classical singular solutions) gives a
monopole-like configuration from an initial classical configuration
which was not monopole-like.}} We will call this a
\textbf{\textit{``quantum monopole''}} to distinguish it
from field configurations which are monopole-like already in
the classical theory.

\section{Energy density}

The divergence of the fields of the classical solution
given by Eqs. \eqref{s1-5} - \eqref{s1-7} leads to a diverging
energy density for the solution, and thus an infinite
total field energy. The energy density $\varepsilon$ 
of the quantized solution is
\begin{equation}\label{en-0}
  \varepsilon \propto (E^a_\mu)^2 + (H^a_\mu)^2
   \propto \left( \frac{rg' - g}{r^2}\right)^2 +
   \frac{2\langle f^2\rangle g^2}{r^4} +
   \frac{2\langle {f'}^2\rangle}{r^2} +
   \frac{\langle (f^2 - 1)^2 \rangle}{r^4}
\end{equation}
The first two terms on the right hand side of Eq.
\eqref{en-0}, which involve the ``classical''
ansatz function $g(r)$, go to zero faster than
$1/r^4$ as $r \rightarrow \infty$ due to the form
of $g(r)$ in Eq. \eqref{av-9}. Thus the leading behavior
of $\varepsilon$ is given by the last two terms in
Eq. \eqref{en-0} which have only the ``quantum''
ansatz function $f(r)$. To calculate $\langle {f'}^2 \rangle$
let us consider
\begin{equation}
  r^2_1 \frac{d}{dr_1}
  \left\langle 
  f'(r_2) f'(r_1)
  \right\rangle = \left\langle f'(r_2) f^3(r_1) \right\rangle - 
  \left\langle f'(r_2) f(r_1) \right\rangle 
  \left( 1 + g^2(r_1) \right)
\label{en-1}
\end{equation}
In the limit $r' \rightarrow r$ we have 
\begin{equation}
  r^2 \left\langle f'(r) f''(r) \right\rangle = 
  \frac{r^2}{2} \left\langle {f'}^2(r) \right\rangle ' = 
  \frac{1}{4} \left\langle f^4(r) \right\rangle ' - 
  \frac{1}{2} \left\langle f^2(r) \right\rangle ' 
  \left( 1 + g^2(r) \right).
\label{en-2}
\end{equation}
From Eqs. \eqref{av-4} and \eqref{av-6} we see that
$\langle {f'}^2 \rangle ' = 0$ which implies
$\langle {f'}^2 \rangle  = const.$
Thus the third term in \eqref{en-0} gives the leading
asymptotic behavior as $r \rightarrow \infty$ to be
\begin{equation}
  \varepsilon \approx \frac{const}{r^2}
  \label{en-3}
\end{equation}
and the total energy of this ``quantum monopole'' (excitation) 
is infinite. This fact indicates that our approximation \eqref{av-2} 
is good only for $\langle f^n(r) \rangle$ calculations but not for 
the derivative $\langle {f'}^2(r) \rangle$.

\section{Conclusions}

\begin{figure}[htb]
\begin{center}
\fbox{
\includegraphics[height=5cm,width=8cm]{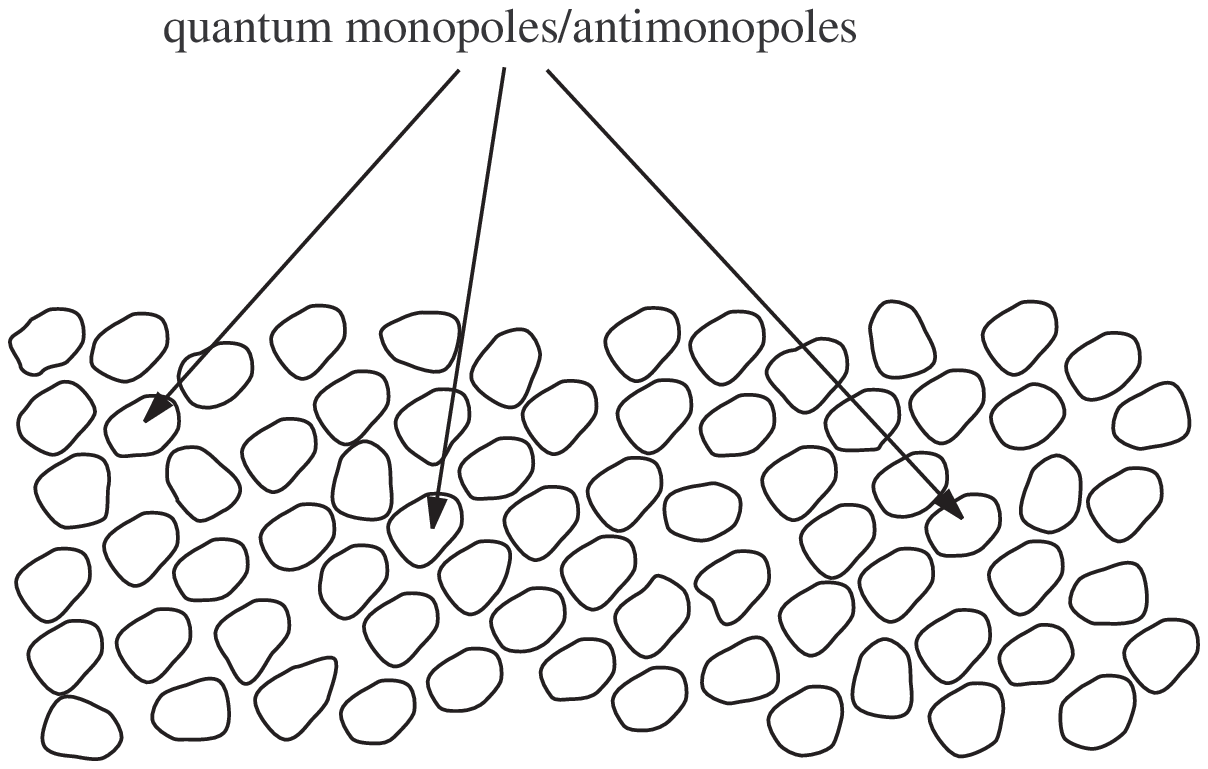}}
\caption{QCD vacuum $\approx$ stochastic gas of quantum
monopoles/antimonopoles.}
\label{fig3}
\end{center}
\end{figure}

Starting from an infinite energy, classical solution to the
SU(2) Yang-Mills field equations we found that the bad
asymptotic behavior of this solution was favorably modified by a
variation of the quantization method proposed by Heisenberg
to deal with strongly coupled, nonlinear field theories. In addition,
although the original classical solution was not monopole-like,
it was found that the quantized solution was monopole-like.
One possible application of this
is to the dual-superconductor picture of the QCD vacuum. In this
picture one models the QCD vacuum as a stochastic gas of
appearing/disappearing monopoles and antimonopoles as in
Fig.\ref{fig3}. These monopole/antimonopole fluctuations can
form pairs (analogous to Cooper pairs in real superconductors)
which can Bose condense leading to a dual Meissner effect
\cite{hooft2} \cite{mand}
expelling color electric flux from the QCD vacuum,
expect in narrow flux tubes which connect and confine the
quarks. Lattice calculations confirm such a
model \cite{suzuki} : monopoles appear to play a
major role in the QCD lattice gauge path integral.
Based on the results of the present paper it may be that : 
(a) the monopoles which are considered in the dual
superconductor QCD vacuum picture and (b) the monopoles in 
lattice simulations whose location is determined by the Abelian 
Projection procedure should be the ``quantum'' monopoles discussed 
here.

\section{Acknowledgment}

VD is grateful for Viktor Gurovich for the fruitful discussion, 
ISTC grant KR-814 for the financial support and Alexander von Humboldt 
Foundation for the support of this work.


\begin{thebibliography}{}

\bibitem{dirac} P.A.M. Dirac, Proc. Roy. Soc., {\bf A133}, 60
(1931); P.A.M. Dirac, Phys. Rev. {\bf 74}, 817 (1949)

\bibitem{thooft} G. 't Hooft, Nucl. Phys. {\bf B79}, 276 (1974);
A.M. Polyakov, JETP Lett. {\bf 20}, 194 (1974)

\bibitem{pol} 
M.~I.~Polikarpov,
``Recent results on the Abelian projection of lattice gluodynamics,''
Nucl.\ Phys.\ Proc.\ Suppl.\  {\bf 53}, 134 (1997), 
hep-lat/9609020.

\bibitem{hs1}
W. Heisenberg, Nachr. Akad. Wiss. G{\"o}ttingen,
N8, 111 (1953);
W. Heisenberg, Nachr. Akad. Wiss. G{\"o}ttingen;
W. Heisenberg, Zs. Naturforsch., {\bf 9a}, 292 (1954); W. Heisenberg,
F. Kortel and H. M{\"u}tter, Zs. Naturforsch., {\bf 10a}, 425 (1955);
W. Heisenberg, Zs. f{\"u}r Phys., {\bf 144}, 1 (1956); P. Askali and
W. Heisenberg, Zs. Naturforsc., {\bf 12a}, 177 (1957); W. Heisenberg,
Nucl. Phys., {\bf 4}, 532 (1957); W. Heisenberg, Rev. Mod. Phys.,
{\bf 29}, 269 (1957)

\bibitem{hs3}
W. Heisenberg, {\it Introduction to the Unified Field
Theory of Elementary Particles}., Max-Planck-Institute f{\"u}r Physik
und Astrophysik, Interscience Publishers London, New York, Sydney,
1966

\bibitem{wu} T.T. Wu and C.N. Yang {\it Properties of Matter Under
Unusual Conditions}, ed. H. Mark and S. Fernbach (Interscience, New
York 1968)

\bibitem{eich} E. Eichten, {\it et. al.}, Phys. Rev. {\bf D17},
3090 (1978)

\bibitem{yos} D. Singleton and A. Yoshida, Int. J. Mod. Phys.
{\bf A12}, 4823 (1997)

\bibitem{thooft2} G. t'Hooft, Nucl. Phys., \textbf{B190}, 455 (1981); 
Proccedings of the Workshop \textit{``Confinement, Duality and 
Non-perturbative Aspects of QCD''}, Cambridge (UK), 24 June - 2 July, 
1997. 

\bibitem{kronfeld} 
A.S. Kronfeld, M.L. Laursen and U.J. Wiese, Phys. Lett. \textbf{190B}, 
516(1987); 
A.S. Kronfeld, G. Schierholz and U.J. Wiese, Nucl. Phys. \textbf{B293}, 
461 (1987). 

\bibitem{kondo}
T.~Shinohara, T.~Imai and K.~I.~Kondo,
``The most general and renormalizable maximal Abelian gauge,'', 
hep-th/0105268.

\bibitem{pol2}
M.~N.~Chernodub and M.~I.~Polikarpov, 
``Abelian projections and monopoles,'', 
hep-th/9710205.

\bibitem{dzh4} V. Dzhunushaliev, Phys. Rev, \textbf{B64}, 024522 
(2001). 

\bibitem{dzh2}
V. Dzhunushaliev and D. Singleton,
Int. J. Theor. Phys, \textbf{38}, 887(1999); hep-th/9912194.

\bibitem{dzh3} V. Dzhunushaliev, Phys. Lett. {\bf B498}, 218
(2001); hep-th/0010185

\bibitem{callan} C.G. Callan, R. Dashen and D.J. Gross, Phys. Lett.
{\bf B66}, 375 (1977); Phys Rev. {\bf D17}, 2717 (1978);
Phys. Rev. {\bf D19}, 1826 (1979)

\bibitem{sn} J.V. Steele and J.W. Negele, Phys. Rev. Lett.,
{\bf 85}, 4207 (2000)

\bibitem{negele} J.W. Negele, ``Instanton and Meron Physics in
Lattice QCD'', hep-lat/0007027; J.V. Steele, ``Can Merons
Describe Confinement?'', hep-lat/0007030

\bibitem{hooft2}
G. t'Hooft, in Proc. Europ. Phys. Soc. Conf. on High Energy
Physics (1975), p.1225.

\bibitem{mand}
S. Mandelstam, Phys. Rev. {\bf D19}, 2391 (1979).

\bibitem{suzuki}
T. Suzuki and I. Yotsuyanagi, Phys. Rev. {\bf D42},
4257(1990);
J.D. Stack, S.D. Neiman and R.J. Wensley, Phys. Rev. {\bf D50},
3399(1994);
M.N. Chernodub and M.I. Polikarpov, ``Abelian projections
and monopoles'', hep-th/9710205.

\end{thebibliography}
\end{document}